\documentclass{aa}

\usepackage{graphics,psfig,amssymb,float}

\newcommand{\ltap}{\mathrel{\hbox{\rlap{\lower.55ex \hbox {$\sim$}}

                   \kern-.3em \raise.4ex \hbox{$<$}}}}

\begin{document}


\title{Photometric variability in the open cluster M\,67}
\subtitle{I. Cluster members detected in X-rays}

\titlerunning{Photometric variability in M\,67 - I. X-ray sources}

\author{Maureen van den Berg\inst{1}\thanks{\emph{Present adress:} 
Osservatorio Astronomico di Brera, Via E. Bianchi 46, 23807 Merate (LC), 
Italy} \and Keivan Stassun\inst{2} \and Frank Verbunt\inst{1} \and Robert  
D. Mathieu\inst{2}}
\authorrunning{Maureen van den Berg et al.}

\offprints{Maureen van den Berg}

\mail{vdberg@merate.mi.astro.it}

\abstract{We study photometric variability among the optical
counterparts of X-ray sources in the old open cluster M\,67.  The two
puzzling binaries below the giant branch are both variables: for
S\,1113 the photometric period is compatible with the orbital period,
S\,1063 either varies on a period longer than the orbital period, or
does not vary periodically. For the spectroscopic binaries S\,999,
S\,1070 and S\,1077 the photometric and orbital periods are similar.
Another new periodic variable is the main-sequence star S\,1112, not
known to be a binary. An increase of the photometric period in the
W\,UMa system S\,1282 (AH\,Cnc) is in agreement with a previously
reported trend. Six of the eight variables we detected are binaries
with orbital periods of 10 days or less and equal photometric and
orbital periods. This confirms the interpretation that their X-ray
emission arises in the coronae of tidally locked magnetically
active stars. No variability was found for the binaries with orbital
periods longer than 40 days; their X-ray emission remains to be
explained.  
}

\institute{   Astronomical Institute, Utrecht University,
              P.O.Box 80000, 3508 TA Utrecht, The Netherlands
         \and Department of Astronomy, University of
              Wisconsin-Madison, 
              475 N Charter St, Madison WI 53706 USA}

\date{Received date / Accepted date}   

\maketitle

\section{Introduction}
Twenty five members of the old open cluster M\,67 have been detected
in X-rays (Belloni et al. 1998). At the age of M\,67 (4 Gyr, Pols
et al. 1998) the rotation of single stars is too slow to generate
detectable X-rays.  Therefore, the X-ray emission of many M\,67
sources probably arises in interacting binaries. Indeed, one source is
known to be a cataclysmic variable. Nine sources are binaries with
orbital periods of 10 days or less, presumably RS\,CVn type systems,
whose X-rays are due to the coronae of magnetically active stars
forced to corotate by tidal interaction (see Table~\ref{Xlist}).
However, not all X-ray sources are binaries, e.g. one source is 
a hot white dwarf, and some others are stars which do not show
signs of binarity.

There are five peculiar binaries (S\,1040, S\,1063, S\,1072, S\,1082
and S\,1113) whose evolutionary statuses we currently do not
understand. They are found in the colour-magnitude diagram in
locations which cannot be reproduced by combining the light from any
two members on the main sequence, subgiant and/or giant branches. Two
of them have long orbital periods which exclude strong tidal
interaction. A sixth binary (S\,1237) with a long orbital period lies
to the blue of the giant branch which can be explained by the
superposition of the light of a giant and a turnoff star.

A spectroscopic study of these peculiar systems was presented in van
den Berg et al. (1999); see their Fig.~1 for the position of these
systems in the colour-magnitude diagram of M\,67. Here we report our
photometric study of these sources and of other X-ray sources that
happen to be in the same fields of view. The photometry of S\,1082 is
published separately (van den Berg et al. 2001). The observations and
analyses are described in Sect.~\ref{obs}. Results are presented in
Sect.~\ref{results}, followed by the interpretation and discussion --
including comparison with earlier work -- in Sect.~\ref{disc}.
Sect.~\ref{concl} summarizes our conclusions. The variability of stars
not detected in X-rays but included in our observations will be the
subject of Paper II (Stassun et al. 2001).

\nocite{bellea} \nocite{vdbergea} \nocite{stasead} \nocite{polsea}
\nocite{san} \nocite{montea} \nocite{mathlathea}

\section{Data and analysis} \label{obs}

\subsection{Observations}

\begin{figure*}
      \resizebox{\hsize}{!}{\includegraphics{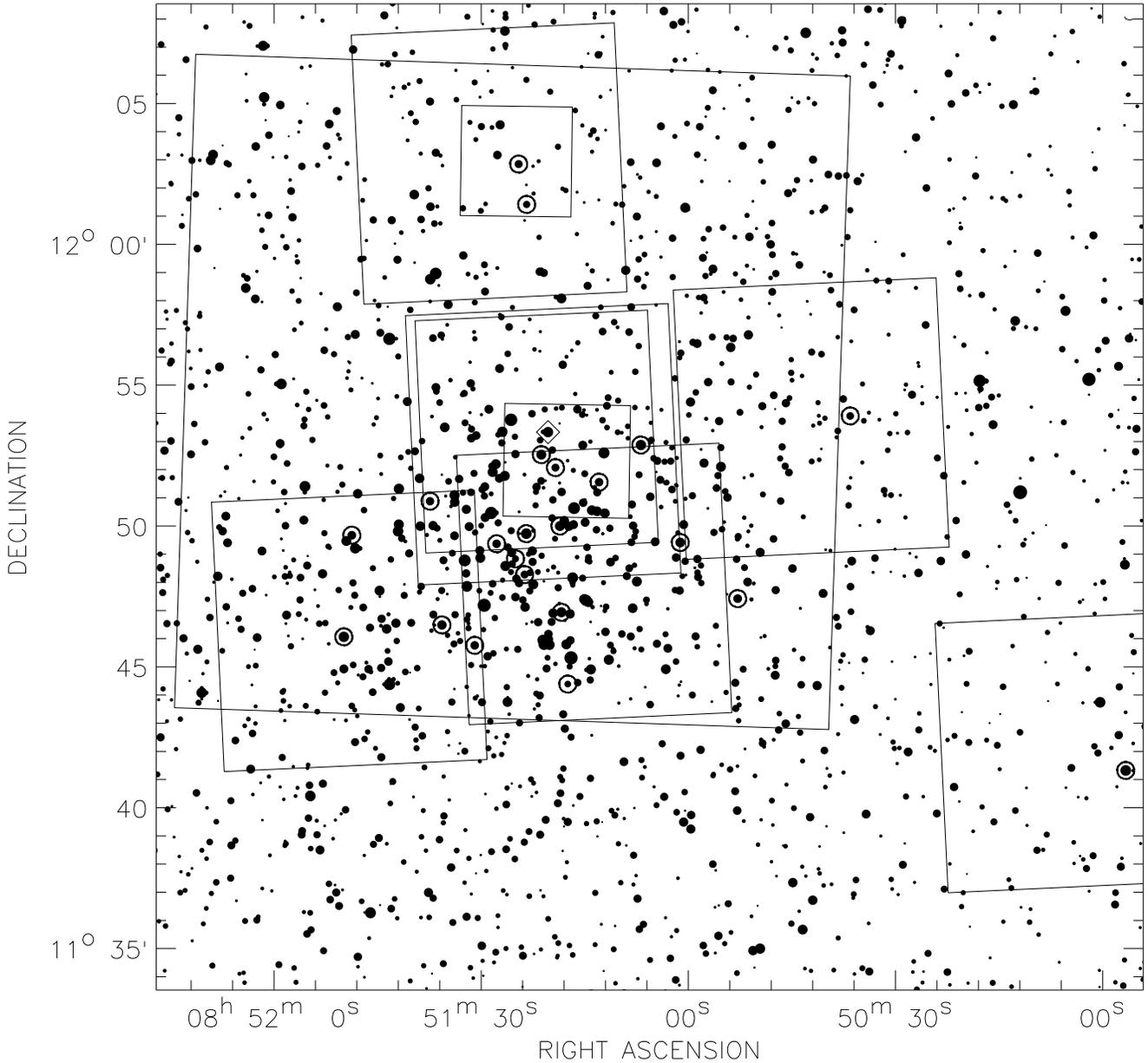}}
      \caption{A 35\arcmin x 35\arcmin\ region of M\,67 centered on
      the star S\,783. The coordinates of the sources in this field
      are taken from the USNO-A1.0 catalogue (epoch J2000). The areas
      monitored in the five observation runs are indicated by the
      squares (see also Table~\ref{log}). The open circles mark the
      optical counterparts to the X-ray sources that are discussed in
      this paper; the diamond marks S\,1082 that we discuss elsewhere
      (van den Berg et al. 2001).  \label{map}}
\end{figure*}

$U$, $B$, $V$, $I$ and Gunn\,$i$ photometry was obtained with the 0.90m
telescope at Kitt Peak, the 0.91m ESO-Dutch Telescope at La Silla and
the 1m Jacobus Kapteyn Telescope on La Palma. Combined, the five
observation runs span a period of two years (see Table~\ref{log}).
Fig.~\ref{map} shows the location of the observed fields. During the
first run weather conditions were good with a typical seeing of 0\farcs9 to
1\farcs4. The observations were made during and around full moon but
moon illumination was not a problem. Weather conditions during runs 2
and 3 were good, with a typical seeing of 1\farcs6. During run 4 the
typical seeing was 1\farcs5 while the quality of some images was
affected by the brightness of the nearby moon. The same, at a seeing
between 1\farcs5 and 3\arcsec, is true for the last run, that in addition
was troubled by partial cloudiness. This is reflected in the relatively
large errors of the last two runs.

Every X-ray source of Belloni et al. (1998) was monitored in at least
one run, except for the faint cataclysmic variable EU\,Cnc and the hot
white dwarf. The main purpose of run 1 was to monitor variability of
S\,1113 and S\,1063, of run 2 to monitor S\,1113 and of runs 3 to 5 to
monitor S\,1082.  This means that exposure times were chosen to
optimize the measurements of these stars. During run 5, five
additional fields containing X-ray sources were observed in $B$ and $V$
once or twice per night to search for obvious signs of variability. As
S\,364 and S\,1237 are very bright stars (see Table~\ref{Xlist}) the
quality of the images of fainter stars in these fields are poor. This
affects the quality of the light curves of the X-ray sources in the
field of view of S\,1237 (i.e. S\,1242, S\,1270 and S\,1282).

\begin{table*}
\caption{Log of the observations. From left to right: number and dates
of the
observation run; telescope; Sanders number of the star near the centre of
the field; 
field of view; filters; typical exposure time in seconds for each
filter; 
the minimum and maximum period $P_{{\rm min}}$ and $P_{{\rm max}}$ used
to 
compute periodograms for the specified run.}
\begin{tabular}{l@{\hspace{0.2cm}}lll@{\hspace{0.2cm}}lllr@{\hspace{0.2cm}}r}
Run & Dates                   & Telescope      & Centre           
  & FOV  & Filters       & $t_{{\rm exp}}$ (s) & $P_{{\rm min}}$ &
$P_{{\rm max}}$ \\
  & & & & & & & &  \\
1   & Jan 4-16 1998     & 0.90m KPNO & S\,1084  &  23\arcmin
x 23\arcmin & 
$B\,V\,I$          & 90~60~60 & 2.4 hr  & 10 d  \\
2   & Feb 12,14-17,21,24-28  & 0.91m ESO Dutch & S\,1113    &  3\farcm8
x 3\farcm8   & 
$U\,B\,V$\,Gunn\,$i$ & 300\,120\,120\,120 & 5 hr & 25 d \\
    & 1998, Mar 1-3,6,8  1998  & & & & &  & &  \\
3   & Feb 2-19 1999           & 0.91m ESO Dutch & S\,1068$^{*}$ &
3\farcm8 x 3\farcm8 & 
$U\,B\,V$\,Gunn\,$i$ & 360~100~50~30 & 30 mn & 18 d \\
4   & Dec 25, 26 1999         & 1m ING JKT     & S\,1082 & 10\arcmin x
10\arcmin & 
$B\,V$ & 75~30 & 12 mn & 1 d \\
5   & Feb 13-16, 20 2000      & 1m ING JKT     & S\,1082   & 10\arcmin x
10\arcmin & 
$U\,B\,V\,I$ & 350~30~15~8 &  30 mn & 8 d \\
    &                         &                & S\,364    & 10\arcmin x
10\arcmin & 
$B\,V$  & 15~4 & 4 hr & 8 d \\
    &                         &                & S\,628    & 10\arcmin x
10\arcmin & 
$B\,V$  & 150~80 & 4 hr & 8 d \\
    &                         &                & S\,1013   & 10\arcmin x
10\arcmin & 
$B\,V$  & 20~8 & 4 hr & 8 d \\
    &                         &                & S\,1113   & 10\arcmin x
10\arcmin & 
$B\,V$  & 200~100 & 4 hr & 8 d \\
    &                         &                & S\,1237   & 10\arcmin x
10\arcmin & 
$B\,V$  & 20~10 & 4 hr & 8 d \\
    & & & & & & & & 
\end{tabular}

$^*$ S\,1063 starting from February 8
\label{log}
\end{table*}

\begin{table*}
\caption{ Properties of the X-ray sources of Belloni et al. (1998)
discussed in this paper. From left to right: Sanders number (Sanders
1977); X-ray source number from Belloni et al. (1998); $V$ magnitude and
$B-V$ colour (Montgomery et al. 1993); for spectroscopic binaries:
orbital period $P_{\rm orb}$ in days and eccentricity $e$; observation
runs during which the star was observed; a variability indicator {\em
n} ({\em y}) if the probability that the source is constant is larger
(smaller) than 0.3\%; photometric period $P_{\rm phot}$ in days with
the run number(s) and the filter for which the period was detected in
square brackets; maximum phaseshift used to estimate the error in the
period; logarithm of the false alarm probability of the period
detection; remarks and references for information on the spectroscopic
binaries (1=Mathieu et al. 2002, in preparation, 2=Mathieu et
al. 1990, 3=preliminary solution by Latham et al., private communication).}

\begin{tabular}{r@{\hspace{0.25cm}}r@{\hspace{0.25cm}}r@{\hspace{0.25cm}}r@{\hspace{0.25cm}}r@{\hspace{0.25cm}}r@{\hspace{0.25cm}}r@{\hspace{0.25cm}}r@{\hspace{0.25cm}}rllr@{\hspace{0.25cm}}rl}
   S & X &   $V$  &   $B-V$ & $P_{\rm orb}$ (d) & $e$          & run & var &
\multicolumn{2}{c}{$P_{\rm phot}$ (d)}   & $d\phi_{\rm max}$ & log FAP & 
remarks \\
     &    &       &       &           &            &     &     &  &  & 
&  &         \\
 364 & 19 &  9.84 &  1.36 &           &            & 5   & n   &  &  & 
&  &         \\
 628 & 35 & 14.47 &  0.76 &           &            & 5   & n   &  &  & 
&  &         \\
 760 & 49 & 13.29 &  0.57 & 954(12)   & 0.43(9)    & 1   & n   &  &  & 
&  & 3       \\
 775 & 44 & 12.69 &  0.63 &           &            & 1,5 & n   &  &  & 
&  &         \\
 972 & 17 & 15.37 &  0.89 & 1.166412(2) & 0.009(6) & 1,5 & n   &  &  & 
&  & 3       \\
 999 & 13 & 12.60 &  0.78 & 10.05525(19)  & 0.  & 1,5 & y   &
9.2(2.0)$^a$ & [1;$B$] & 0.3 & -2.62 & 2 \\
1019 & 11 & 14.34 &  0.81 & 1.360217(18) & 0.023(13) & 1,5 & n   &  &  &
 & &  3        \\
1027 & 46 & 13.24 &  0.60 &           &            & 1,4,5 & n &  &  & 
& &       \\
1036 & 45 & 12.78 &  0.49 &     &            & 1,4,5 & y &
0.44144(1)$^{a}$ & [all;$B$] & 0.05 & -18.6 &  EV\,Cnc  \\
     &  &  &  &  &   &       &  & 0.44144(1)$^{a}$ & [all;$V$] & 0.05    &
-17.0 &  \\
     &  &  &  &  &   &       &  & 0.44144(1)$^{a}$ & [1+5;$I$] & 0.05    &
-6.4 &  \\
1040 & 10 & 11.52 &  0.87 & 42.8271(22)   & 0.  & 1,4,5 & n &  &  &  & 
&         \\
1045 & 41 & 12.54 &  0.59 & 7.64521(11)   & 0.   & 1,4,5 & n &  &  &  & 
&         \\
1063 &  8 & 13.79 &  1.05 & 18.396(5)    & 0.206(14)  & 1,3,4,5 & y &
(17-18$^{b}$) & [1;$U$], & &   & 1        \\
     &  &  &  &  &   &       &  & & \,[1+3;$B,V,I$] &     & &\\
1070 & 38 & 13.90 &  0.63 & 2.66059(8)   & 0.  & 1,3,4,5 & y & 2.6(1) &
[3;$B$] & 0.3 & -17.6 & 3 \\
     &  &  &  &  &   &       &  & 2.6(1) & [3;$V$]  & 0.3 & -31.5 &  \\
1072 & 37 & 11.32 &  0.61 & 1495(16)     & 0.32(7)    & 1,3,4,5 & n &  &
 &  &     & 2 \\
1077$^c$ &  7 & 12.60 &  0.64 & 1.358766(8) & 0.095(33) & 1,3,5   & y &
1.42(9) & [3;$B$] & 0.3 & -3.3 & 3 \\
     &  &  &  &  &   &       &  & 1.44(9)$^{a}$ & [3;$V$] & 0.3 & -7.2 &  \\
     &  &  &  &  &   &       &  & 1.38(9) & [3;$I$] & 0.3     & -5.5  &  \\
1112 & 28 & 14.98 &  0.78 &           &            & 1,2,5 & y   &
2.7(2) & [1;$B$] & 0.3 & -3.0 &  \\
     &    &       &       &           &            &       &     &
2.65(3) & [1+2;$V$] & 0.3 & -7.8 & \\
1113 & 26 & 13.77 &  1.01 & 2.823105(14)  & 0.022(10)         & 1,2,5 &
y   & 2.84(8) & [2;$U$] & 0.1 & -3.9  & 1, AG\,Cnc \\
     &    &       &       &           &            &       &     &
2.834(1) & [all;$B$] & 0.1 & -14.5 & \\
     &    &       &       &           &            &       &     &
2.833(1) & [all;$V$] & 0.1 &  -14.9& \\
     &    &       &       &           &            &       &     &
2.84(3) & [1+2;$I$] & 0.1 & -11.2  & \\ 
1234$^c$ & 53 & 12.65 &  0.57 & 4.35563(25)   & 0.  & 1     & n   &  &  
    &   &  & 2\\
1237 & 52 & 10.78 &  0.94 & 697.8(7)     & 0.105(15)  & 5     & n   & & 
&     &     & 2 \\
1242 & 50 & 12.72 &  0.68 & 31.7797(27)   & 0.664(18)  & 1,5   & n   & &
 &     &  & 2 \\
1270 & 43 & 12.73 &  0.58 &           &            & 1,5   & n   &  &  &
    &  & \\
1282 & 40 & 13.33 &  0.56 &   &     & 1,4,5 & y   & 0.360452(8)$^{a}$  &
[all;$B$] & 0.05 & -19.2 & AH\,Cnc \\
     &  &  &  &  &   &       &  &  0.360452(8)$^{a}$  & [all;$V$] & 0.05  
  & -21.9 &  \\
\end{tabular}

$^a$ period does not correspond to the highest peak in the computed
periodogram \\
$^b$ folded light curves do not look convincing\\ 
$^c$ this system is a triple system; the period listed is for the
inner binary
\label{Xlist}
\end{table*}

\subsection{Data reduction and light curve solution} 

Standard IRAF\footnote{IRAF is distributed by the National Optical
Astronomy Observatories, which are operated by the Association of
Universities for Research in Astronomy, Inc., under cooperative
agreement with the National Science Foundation} routines were
used to remove the bias signal and flatfield the images. Aperture
photometry for all the stars was done with the {\sc daophot.phot}
task. For each individual run, source counts were extracted within a
fixed radius with a value depending on the seeing conditions. The
stars in M\,67 are separated well enough to avoid problems of
crowding.

The light curve solution was computed with the algorithm of ensemble
photometry as described by Honeycutt (1992).  In this method, the
magnitudes of all the stars on every frame are used to create an
ensemble average with respect to which the brightness variations are
defined.  Frames that have a large offset from this average (e.g. due
to bad seeing) show up as deviant observations and can be
excluded. For a given star, errors were assigned to the data points by
estimating the typical spread in the light curves of stars of similar
magnitude. In most cases the formal errors from the {\sc phot} task
are negligible; if not, we used this error instead. The different
datasets were analyzed individually.

As exposure times were chosen to optimize the measurements of the
X-ray sources in each field, the photometric precision as a function
of stellar brightness varies from one run to the next. Generally
speaking, the photometric precision of the brightest (unsaturated)
stars in our exposures is flat-field limited to 5--10 mmag. This
precision level typically holds for stars up to 2--2.5 mag fainter
than the brightest sources, and then becomes photon-noise limited and
degrades for still fainter stars. The best overall precision was
achieved on our Kitt Peak frames (run 1; Table~\ref{log}), for which
the brightest stars ($B\approx 12$, $V \approx 12$, $I \approx 11.5$)
have $\sigma_{\rm mag} = $ 0.007, 0.005, 0.005 in $B$, $V$, and $I$,
respectively.  The precision begins to degrade at around 14th mag. For
the faintest sources, at about 18.5 mag, the precision is 0.05--0.1
mag. We refer the reader to Paper II for a full discussion of the
photometric precision in our observations.

A simple zero-point shift is applied to the measurements in each
filter to roughly place the instrumental magnitudes on an absolute
scale as described in Paper II. The light and colour curves that are
presented in Fig.~\ref{ahcnc}-\ref{noper} show the variations with
respect to the mean magnitude and colour as listed in Table~3 of Paper II.

\nocite{honn}

\subsection{Search for variability} \label{varana}

Our search for variability is a two-step process.  First we perform a
$\chi^2$-test on the individual light curves for each filter for each
run, to calculate the probability that the light curves of the X-ray
sources are compatible with being constant.  As the intrisic
properties of the variability need not be the same in light curves of
different runs (in particular for brightness variations due to spots),
we consider each light curve separately. To remove accidental
outliers, the minimum and maximum data points are excluded. A star is
listed as a variable in column ``var'' of Table~\ref{Xlist} if the
$\chi^2$-test predicts for any of its light curves that the
probability for it being constant is smaller than 0.3\%. Eight stars
are thus marked as probable variables. For six of these stars the
variability is not detected in every light curve. In most cases we can
ascribe this to differences in sensitivity between runs or between
different filters of a certain run, or to different durations of
runs. For S\,1063, S\,1070 and S\,1077 it seems that the variability
itself has changed as discussed below.

We next perform a Lomb-Scargle time-series analysis (Scargle 1982) to
search for periodicity in the light curves. In cases when multiple
light curves in a given filter are marked as variable, those light
curves are combined for the period search; no distinction was made
between $I$ and Gunn\,$i$. If the resulting period does not produce a
smooth folded light curve, data from different runs are analyzed
separately; this will be indicated for each source in
Sect.~\ref{results}.  A periodogram is computed with 1000 frequencies
between a minimum and maximum period $P_{\rm min}$ and $P_{\rm max}$
corresponding to twice the typical sampling period and the full length
of the longest observation run, respectively (see Table~\ref{log}). We
choose the period of the highest peak in the periodogram as our first
estimate for periodicity in the data. However, as will be discussed
below external information often leads us to immediately neighbouring
peaks of comparable significance.

Photometric periods have been determined previously for the two
contact binaries S\,1036 (Gilliland et al. 1991) and S\,1282 (e.g.
Kurochkin 1979). Therefore, we look for periods in a narrow window
instead of the range limited by $P_{\rm min}$ and $P_{\rm max}$. In
both cases, we find that the power at half the photometric period is
far higher than at the photometric period, due to the symmetry in the
light curve. For S\,1036 we search for periods between 0.215 and
0.225 days, for S\,1282 between 0.175 and 0.185 days. The periodogram
is computed for 5000 points to increase the resolution.

An estimate for the chance detection of a period, i.e. the probability
that the light curve does not have the periodicity indicated by the
highest peak, is expressed by the false alarm probability. In the case
of the Lomb-Scargle periodogram the false alarm probability follows
the expression: 1$-$[1$-$$\exp(-z)]^m$, where $z$ is the height of the
peak and $m$ is the number of independent frequencies. Horne \&\
Baliunas (1986) demonstrated that this number can be smaller than the
number of data points especially in sets of unevenly sampled data.
The value of $m$ is obtained by fitting this expression to a
probability distribution generated by measuring the maximum peak
heights in periodograms of 5000 simulated random datasets with the
same time-sampling and the same spread in the measurements as the
actual light curve.

Photometric periods with a false alarm probability smaller than
1\% are summarized in Table~\ref{Xlist}. These correspond
to the position of the highest peak in the periodogram unless
indicated otherwise.  To estimate the error in the best period we
proceed as follows.  For the correct period the light curves defined by
the first and last observations coincide.  A small change $dP$ in the
period causes a small phaseshift: $d\phi=T/(P+dP)-T/P$, where $T$ is the
timespan of the dataset. For each light curve we estimate, by visual
inspection, a maximum $d\phi_{\rm max}$ for which the light curves do
not split perceptibly.  This corresponds to a maximum acceptable
period change of
\begin{equation}
{dP\over P}={-d\phi_{\rm max} P \over T+d\phi_{\rm max} P}
\end{equation}
The value for $d\phi_{\rm max}$ that we choose is listed in Table~\ref{Xlist}.

\nocite{scar} \nocite{hornbali}

\section{Results} \label{results}

We have divided the sources into periodically (Sect.~\ref{period}) or
non-periodically (Sect.~\ref{noperiod}) varying stars. The
periodically varying stars are two W\,UMa systems, four spectroscopic
binaries and one star not known to be a binary, i.e. S\,1112. 

\begin{figure*}
      \resizebox{\hsize}{!}{\includegraphics{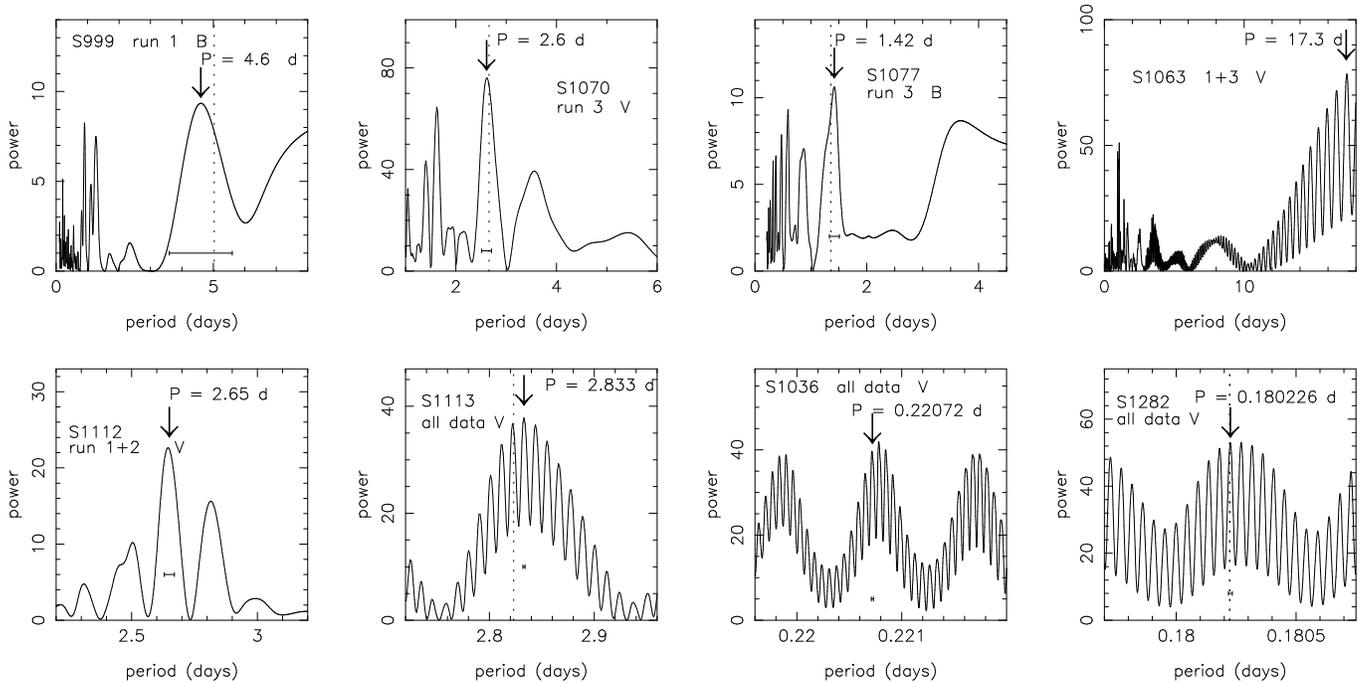}}
\caption{Periodograms for the variable stars. The arrow indicates the
photometric period or half the photometric period listed in
Table~\ref{Xlist}. The dotted line marks the position of the orbital
period for S\,1070, S\,1077 and S\,1113, half the orbital
period for S\,999 and the position of the period as predicted by
Kurochkin (1979) for S\,1282. The horizontal errorbars give our
estimate for the uncertainty in the period.  \label{pgrams}}
\end{figure*}

\subsection{Periodic variables} \label{period}

\subsubsection{W\,UMa systems} 

The W\,UMa light curves are plotted versus {\em photometric} phase
where phase 0 corresponds to the moment of photometric primary
minimum. \\

\noindent
{\bf S\,1282} is the W\,UMa variable AH\,Cnc discovered by
Kurochkin (1960). The periodograms of the total $B$ and $V$ datasets show
peaks with a spacing of $\sim$\,4~10$^{-5}$ and $\sim$\,6~10$^{-4}$ days due
to the two-year and the two-month gaps in our observations, respectively (see
Fig.~\ref{pgrams}).  The highest peaks in the $B$ and $V$ periodograms are
found at 0.180226 and 0.180270 days, respectively, which corresponds
to two neighbouring peaks in the periodogram. We folded the data on
twice those periods but find that the peak at 0.180226 represents best
the true period: when we use the longer period, the deeper primary
minima of the first run fall on top of the secondary minima of the
fourth run.  Thus we conclude that the photometric period is 0.360452
days. $I$ data were only obtained during run 1 and cannot provide an 
equally precise period.

In addition to short-term variations on a time scale of roughly 9 to
10 years,
Kurochkin (1979) finds a secular increase of the period of AH\,Cnc.
His ephemeris for the primary minimum is:
\begin{eqnarray} 
{\rm Min\,I} & = & 2\,441\,740.7166(27) + 0\fd36044098(53)E \nonumber \\
      &   & + 1\fd56(38) 10^{-10} E^2 
\end{eqnarray}
This ephemeris predicts a period during the time of our 
observations between 0\fd3604447 and 0\fd3604479, in agreement
with our result. We note that period changes of similar magnitude have been 
found for other contact binaries.

We observe no significant colour changes in  $V-I$ and $B-V$.
\\
 
\nocite{kuro} \nocite{kuro79}

\begin{figure*}[t]
      \resizebox{\hsize}{!}{\includegraphics{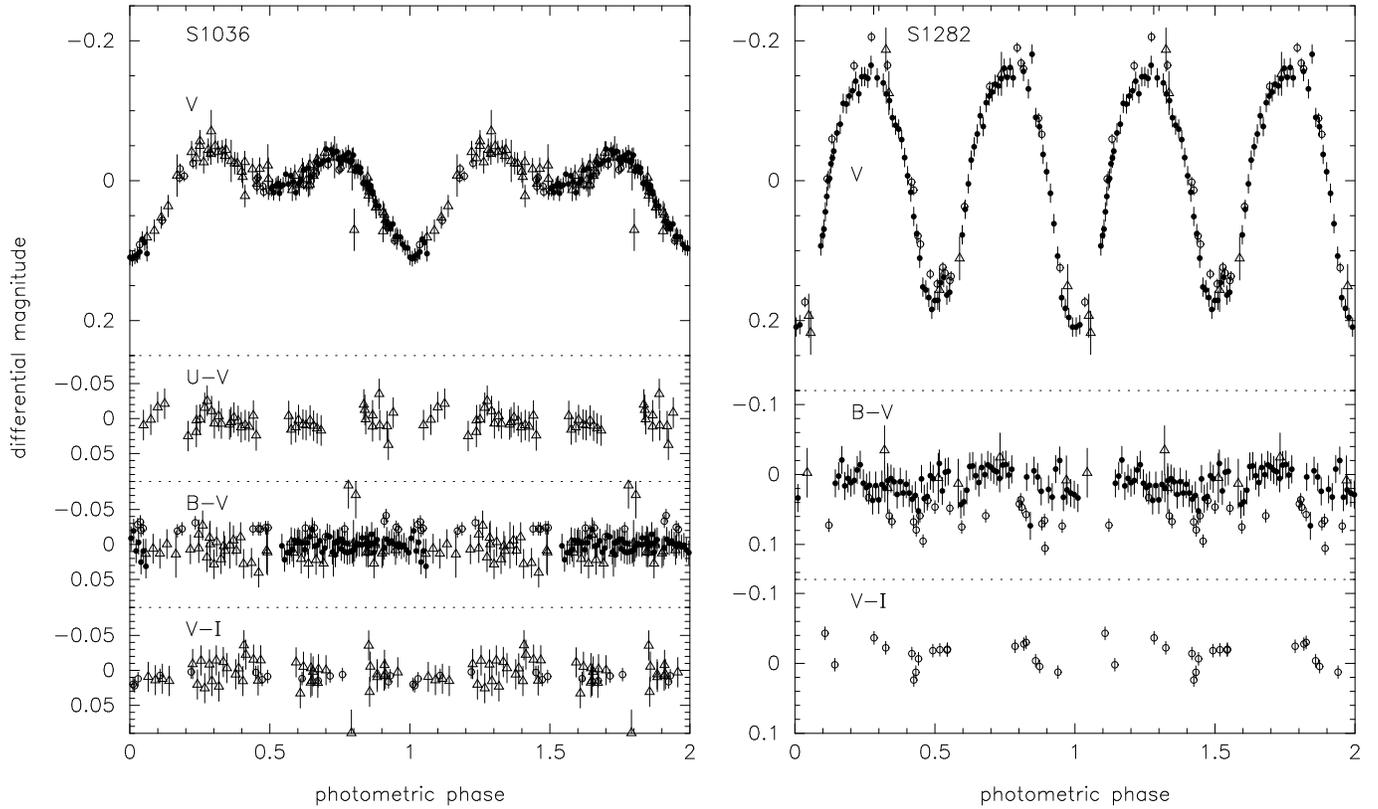}}
      \caption{Light curves of the two contact binaries S\,1036
      (EV\,Cnc) and S\,1282 (AH\,Cnc) folded on the newly derived
      periods (see Table~\ref{Xlist}).  Data from different observing
      runs are marked with different symbols: open circles for run 1,
      filled circles for run 4 and open triangles for run 5.
      \label{ahcnc}}
\end{figure*}

\noindent
{\bf S\,1036}, or EV\,Cnc, was discovered to be a contact binary by
Gilliland et al. (1991) who report a period of 0.44125 days.  The $B$
and $V$ periodograms show fine structure from the two-year and two-month
gaps in the data. In both sets, that include data from run 1, 4 and 5,
we find the same best period of 0.22078 days. When folded on this
period, the light curves show the same effect of interchanging primary
and secondary minima as described for S\,1282. Proper phasing is
obtained with the period of 0.22072 days derived from the peak next to
the highest. The period that corresponds to the highest peak in the
periodogram of the $I$ data is 0.22091 days, but again interchanges
minima. The period of 0.22072 days coincides with  a nearby peak of
similar height. $U$ data were only obtained during run 5 and cannot
provide an equally precise period. In Fig.~\ref{ahcnc} the light
curves are folded on the double period of 0.44144 days.

A light curve folded on the period given by Gilliland et al.\ shows 
that this period cannot be correct for the epoch of our observations. 
Since Gilliland et al. do not specify the error in the period, we 
cannot tell whether the period has changed significantly.

We see no significant changes in any of the colours $U-V$, $B-V$ or $V-I$.

\subsubsection{Spectroscopic binaries}

We note that the light curves of the spectroscopic binaries are gives as
function of {\em orbital} phase where phase 0 corresponds to the
moment of maximum positive radial velocity of the primary star
(primary receding). \\

\noindent
{\bf S\,999} The $B$ and $V$ data of the first run show variability with a
semi-amplitude of $\sim$0.03 mag. Only the period found in the $B$ data
has a false alarm probability smaller than 1\%. The highest peak 
in the periodogram is found at 4.6$\pm$1.0 days (Fig.\ref{pgrams}), but 
this period does not produce a smooth light curve.
We suggest that the peak at 4.6 days is a harmonic of 
photometric variation on or near the orbital period of 10.06 days. Note
that the duration of the first run was 10 nights, so our period search 
does not extend up to the orbital period.
Gilliland et al. (1991) report a period of 9.79 days (no error 
is given) with an amplitude of only 0.013 mag. 
 
\begin{figure*}
      \resizebox{\hsize}{!}{\includegraphics{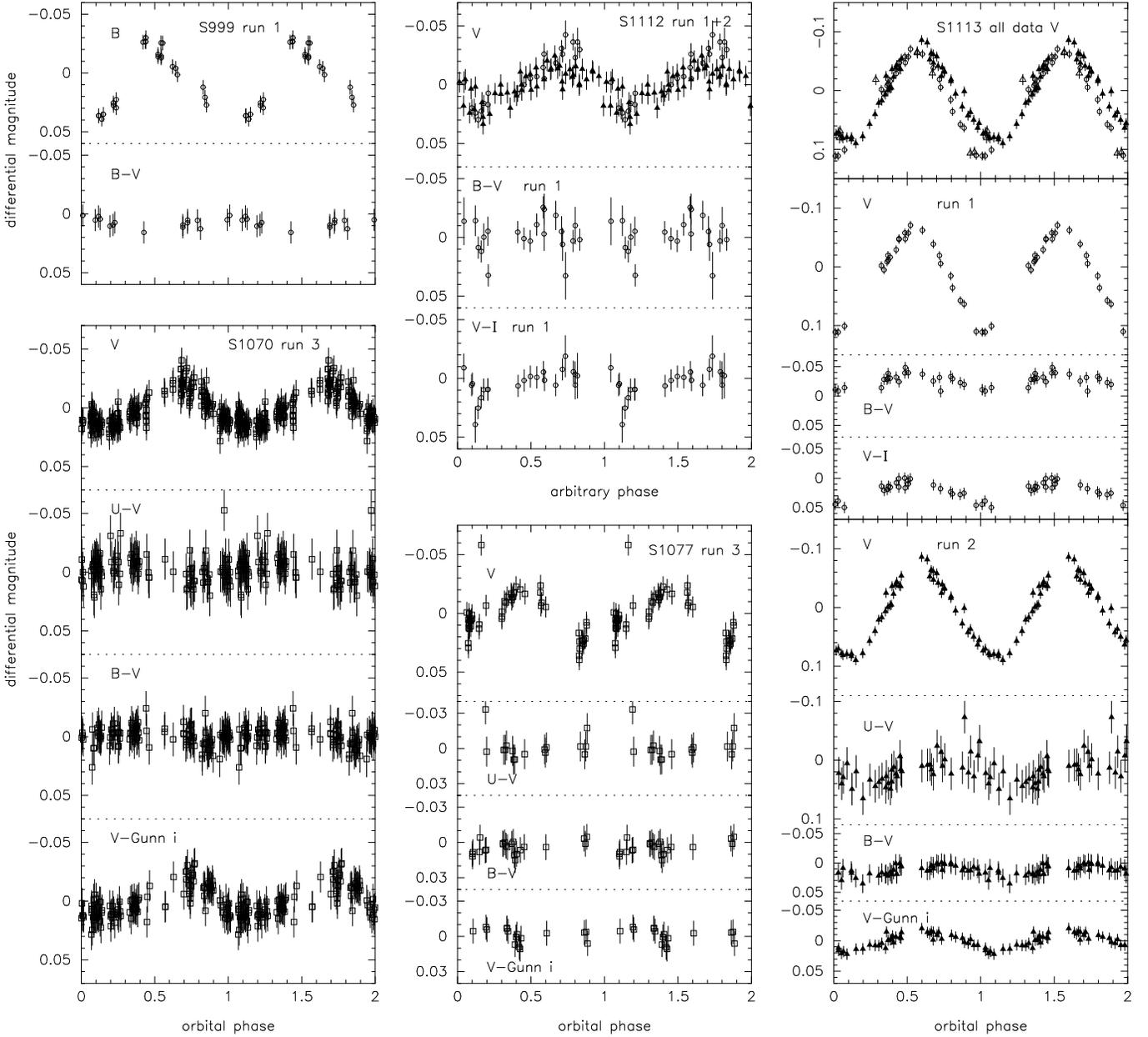}}
      \caption{Light curves of S\,1112 (folded on the photometric
      period) and of S\,999, S\,1070, S\,1077 and S\,1113 (folded on
      the orbital periods; see Table~\ref{Xlist}).  Data from
      different observing runs are marked with different symbols: open
      circles for run 1, filled triangles for run 2, open squares for
      run 3, filled circles for run 4 and open triangles for run 5.
      \label{periods}}
\end{figure*}

In Fig.~\ref{periods} we fold the $B$ and $V$ data on the orbital period.
According to the orbital solution of Mathieu et al. (1990), the
minimum brightness occurs around orbital phase 0.  The $B-V$ colour
does not vary significantly.  \\

\noindent
{\bf S\,1070} A period of 2.6$\pm$0.1 days is detected in the $B$ and $V$ 
light curves of run 3 with a semi-amplitude of the variation of
$\sim$0.03 mag.

The photometric properties of S\,1070 have changed with respect to run
1. If the variations seen in the third run ($\sigma \approx 0.014$ mag
in $V$) were present in the first run, they would have been detected.

In Fig.~\ref{periods} all data of run 3 are folded on the orbital
period which is compatible with the photometric period. The
photometric minimum occurs around orbital phase 0.1-0.2 (ephemeris
from Latham et al., private communication). The $V-$\,Gunn\,$i$ colour curve
shows periodic variations with a semi-amplitude of $\sim$0.03 mag such
that the star becomes bluer as it gets brighter.\\

\noindent
{\bf S\,1077} All light curves of this star are marked as variable
except for the $U$ and $B$ data of run 5, probably due to the reduced
sensitivity of run 5, and the $V$ and $I$ data of run 1. The latter can
point at a real absence of variation as could be the case in S\,1070;
the variations of run 3 ($\sigma \approx$ 0.018 mag in $V$) were not
seen in run 1. In the periodograms of the combined data peaks with a
false alarm probability smaller than 1\% are found near 0.6 and 1.3
days. However, when folded on these periods, the light curves do not
look smooth. Therefore we also analyzed data of the different runs
separately. Only the light curves of run 3, with the highest
precision, look smooth when folded on the periods of about 1.4 days
(see Table~\ref{Xlist}) which have a false alarm probability smaller
than 1\% only in $B$, $V$ and Gunn\,$i$. This period does not correspond to
the highest peak in the $V$ periodogram, which is found at 0.60
days. The semi-amplitude of the variation is small, $\sim$0.03 in $V$.

The photometric period is compatible with the orbital period; we
consider the latter to be the true period for the photometric
variability.  The data of run 3 are folded on the orbital period in
Fig.~\ref{periods}.  The photometric minimum occurs around phase
0.9-0. The colours do not vary significantly. \\

\noindent
{\bf S\,1113} As already noted by Kaluzny \&\ Radczynska (1991) this
star is a photometric variable. In our observations, the $B$ and $V$ data
cover the longest timespan and can therefore provide the most accurate
photometric period.  The periodogram is computed for 25000 periods to
make the period bins smaller than the accuracy of our period
determination (0.001). The periodogram (see Fig.~\ref{pgrams}) again
shows fine structure with a spacing as expected from the two-year gap
between runs 1 and 5. The maximum peak indicates a period of 2.833(1)
days, which is not compatible with the orbital period. The orbital
period corresponds to a neighbouring peak in the periodogram at
2.822(1) days (in $V$; 2.823(1) days in $B$). Given the lack of
significant difference in peak heights, the photometric and orbital
periods may be the same.  Therefore, the light curves in
Fig.~\ref{periods} are folded on the orbital period.  For the $B$ light
curve we find a similar result. The periods found in the $U$ and $I$
band are compatible, but less accurate (see Table~\ref{Xlist}).

Data taken during different runs at the same orbital phase give rise
to scatter in the light curve, as seen in the top panel of
Fig.~\ref{periods}. This could be explained if the amplitude, phase
and/or the period of the variability has changed between the
runs. When analyzed separately, the light curves of run 1 and 2 give a
period of 2.8$\pm$0.1 and 2.83$\pm$0.03 days, respectively; the
uncertainty is too large to detect a period change. Assuming that the
photometric period is the orbital period we conclude that the
amplitude or the phase of the variations has changed, either of which
is possible if the variation is caused by a star spot.

We show separately the light and colour curves from runs 1 and 3 in the
lower panels of Fig.~\ref{periods}. The colour variation is
significant only in $V-I$ in run 1 and $B-V$ and $V-$\,Gunn\,$i$ in run 3, such
that the star becomes bluer as it brightens.  \\

\subsubsection{S\,1112}

The data of the first ($B, V, I$) and second ($V$) runs show variability
but only in the $B$ and $V$ light curves do we find significant periods of
2.7$\pm$0.2 and 2.65$\pm$0.03 days, respectively. In
Fig.~\ref{periods} the data are folded on the latter period. The
amplitude of the variation is again small, only $\sim$0.04 mag. Weak
colour variations are only seen in $V-$\,Gunn\,$i$ and appear to be in phase
with the light curve; the star becomes bluer as it brightens. No
information on binarity from radial-velocity measurements exists for
this star. \\

\subsection{Non-periodic variable: S\,1063} \label{noperiod}

\noindent

\begin{figure}
      \resizebox{\hsize}{!}{\includegraphics{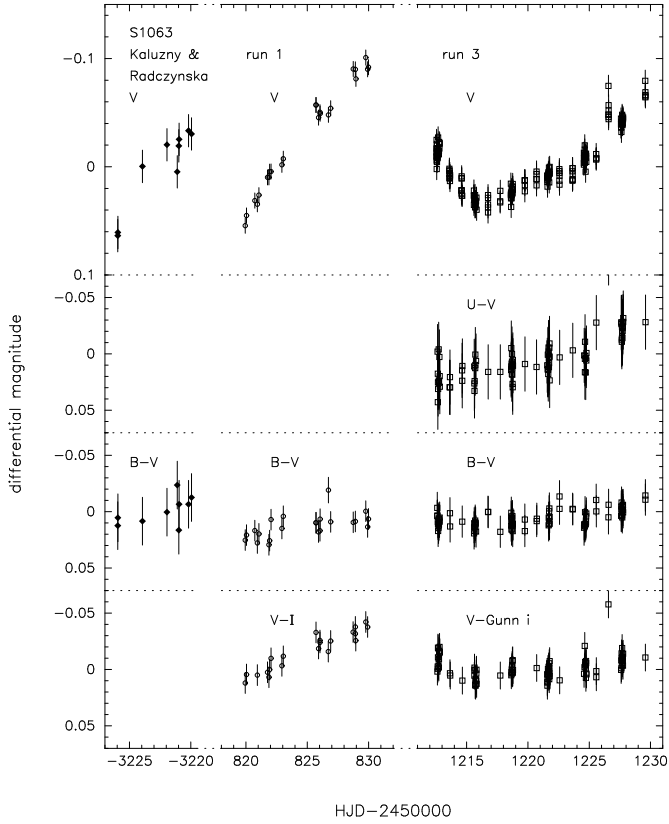}}
      \caption{Light curves and colour curves of S\,1063.  Data from
      different observing runs are marked as in Fig.~\ref{ahcnc}.  The
      data from Kaluzny \&\ Radczynka (1991) of Dec. 9-15 1986 are
      included on the left; the brightness variations are defined with
      respect to the mean magnitude of the lightcurve.
      \label{noper}}
\end{figure}

Photometric variability of this star up to 0.18 mag was inferred by
Racine (1971) from the differences between published values of the
magnitude.  Rajamohan et al. (1988) and Kaluzny \&\ Radczynska (1991)
also noted its variability, but only the latter provide light curves
(for Dec. 9 to 15 1986, see Fig.~\ref{noper}). This binary was
included in all our runs except the second. The light curves of run 1 and
3 clearly show variability (see Fig.~\ref{noper}) on a long time
scale. The variability during run 1 is similar to that observed by
Kaluzny \&\ Radczynska. The longest interval of continuous observation
was eighteen consecutive nights during run 3, which is almost the
length of the orbital period (18.39 days).  Therefore during any one
run we could not have established periodicity on the orbital period.

Data from run 1 and 3 were combined to look for periods up to 18 days.
The highest peaks in the periodogram are found between 17 and 18 days,
all with a false alarm probability smaller than 1\%.  However, the
folded light curves show no convincing periodicity.  Therefore, no
actual period is specified in Table~\ref{Xlist}.

The amplitude of the variation increases towards the blue.
\\

\section{Discussion} \label{disc}

We have studied the optical photometric properties of X-ray sources
in M\,67.  Eight photometric variables, including three new
variables, were found among the twenty two sources that are discussed
in this paper. In all cases the amplitudes of the variations are
small, ranging from 0.03 to 0.4 mag in $V$.

The light curves of the two contact binaries S\,1036 and S\,1282 arise
through partial eclipses and ellipsoidal variations of the tidally
deformed stars. Their X-rays are believed to be emitted by the hot
coronae of the magnetically active components.

The primary and secondary eclipses of contact binaries usually are of
similar depth. This has been interpreted as evidence that both stars
have almost the same temperature, which in turn is evidence for energy
exchange between the two stars in contact.  Unequal depths of the
primary and secondary eclipses then implies different temperatures for
both stars, i.e. poor thermal contact.  The thermal contact can be
suppressed when the system becomes (semi-)detached.  It has been
suggested that such phases of poor thermal contact occur periodically
in contact binaries  (Lucy \&\ Wilson 1979). In view of this
interpretation it may appear surprising that we see unequal eclipses but no
evidence of colour i.e. temperature variations in S\,1036.

S\,1036 is interesting as either an immediate progenitor of a contact binary, 
or because it is in the semi-detached phase of the thermal cycle of a 
contact binary. The upper limit on the colour variations in S\,1036 is
about 0.05 in $B-V$ (Fig.~\ref{ahcnc}). Radial-velocity
measurements are required to determine the evolutionary status of this
system and to convert the upper limit to the colour variations into
an upper limit on temperature difference.  \nocite{lucywils}

The small amplitude of the S\,1036 light curve indicates either a small
inclination or an extreme mass ratio (e.g.\ Rucinski 1997). In a
volume-limited sample of contact binaries, Rucinski (1997) found that
only two among the 98 systems have light curves with unequal
minima. One of those two also has a relatively small amplitude of
variation of about 0.25 mag.

Another feature of W\,UMa light curves associated with unequal eclipses
is that at first quadrature (phase 0.25) the star is brighter than at
second quadrature (phase 0.75). This has been explained with a hot
spot on the secondary, possibly as a result of mass transfer in a
semi-detached system (e.g.  Rucinski 1997). This effect is also
visible in S\,1036. 

Within the two years of our observations we see evidence for
variability of the light curve of S\,1282: the secondary
minimum of run 1 appears to be less deep and flatter than observed in
run 4. A similar variation was seen by Gilliland et al. (1991) who
noted that in their observations of 1988 the secondary eclipse had a
flat shape, while the observations by Whelan et al. (1979) done from
1973 to 1976 showed a rounded secondary minimum. The timescale of
these variations is indicative of the presence of
spots. \nocite{whelea}

The light curves of the periodic variables S\,999, S\,1070, S\,1077
and S\,1113 display only one maximum per cycle. S\,1113 and S\,1070
also show colour variations in phase with the brightness.  For
all four systems there is indication for variability in the light
curves. The amplitude of the variation in S\,999 is different in our
observations and those of Gilliland et al. (1991); in S\,1070 and
S\,1077 there has likely been a change between run 1 and run 3 and in
S\,1113 between run 1 and 2. The short timescale of this variation is
an indication of brightness modulations by spots.  Remarkably, in all
cases the photometric minimum occurs around orbital phase 0. We have no
explanation for this.

All four systems are spectroscopic binaries with photometric periods
compatible with the orbital period. The circular orbital periods,
their X-ray luminosity, and the \ion{Ca}{II} K emission in the case of
S\,999, S\,1077 and S\,1113 (Pasquini \&\ Belloni 1998, van den Berg
et al.  1999) make these stars likely candidates for magnetically
active systems due to one or both stars being tidally locked. This
was already suggested by Belloni et al. (1998) and our light curves
support their interpretation.  \nocite{pasqbell}

The light curve of S\,1112 shows low-amplitude periodic light and
colour variations similar to those seen in these four binaries. This
star has not been monitored for radial-velocity variations but the
X-ray luminosity and the light curve are typical for magnetically
active systems which suggests that S\,1112 is a binary with
an orbital period of about 2.65 days.  \nocite{gillea}

We do not understand the variability that we observe for S\,1063.  The
source shows spectroscopic signatures of magnetic activity (van den
Berg et al.  1999). However, if one of the stars in this binary would
be corotating near periastron, we expect a rotation period of 14.6
days (Eq.~42 of Hut 1981) which is excluded by the observations of run
3. We conclude that either the star does not vary periodically or that
the period of variability is longer than 18 days.  More observations
covering a longer timespan are required to understand the nature of
the variability. Our findings are in contrast with the suggestion by
Kaluzny \&\ Radczynska (1991) that S\,1063, as well as S\,1113, are
highly evolved W\,UMa-type binaries. \nocite{hut}

\section{Conclusion} \label{concl}

Of the twenty two X-ray sources in M\,67 that we discuss, sixteen are
spectroscopic binaries with known orbital periods. Our survey for
optical photometric variables among these X-ray sources has
established eight variables. Seven of these are among the sixteen
binaries, the binary status of the eighth, S\,1112, is not yet
known. In addition, Gilliland et al. (1991) observed periodic optical
variation in three more of the X-ray binaries with amplitudes too low
to be detected by us: S\,1019 (semi-amplitude 0.015 mag), S\,1242
(semi-amplitude 0.0025 mag) and S\,1040 (semi-amplitude 0.012) mag.
Thus ten of the sixteen X-ray binaries in M\,67 are optical variables
at the $\gtrsim$ 0.01 mag level.

\begin{figure}
       \centerline{\psfig{figure=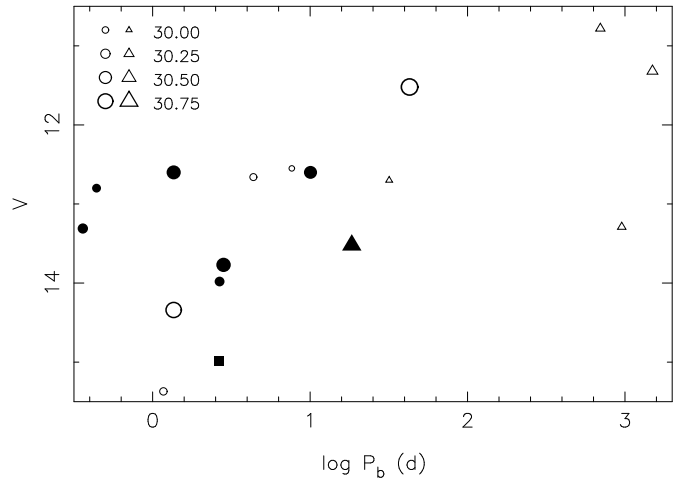,angle=-90,width=\columnwidth,clip=t}
       {\hfill}} \caption{Visual magnitude versus orbital period of
       the M\,67 binaries detected in X-rays. The size of the symbol
       is a measure for the logarithm of the X-ray luminosity (0.1-2.4
       keV, in erg s$^{-1}$) as indicated in the figure. Eccentric
       binaries are indicated with trangles, binaries with
       eccentricities compatible with zero (within the 3$\sigma$
       error) with circles. Filled symbols are systems for which we
       detected photometric variability. S\,1112 is indicated with a
       filled square.  \label{sb}}
\end{figure}

Belloni et al. (1998) have suggested that rapid stellar rotation
resulting from tidal locking results in enhanced magnetic activity and
X-ray emission. Fig.~\ref{sb} shows the visual magnitude versus
orbital period of the spectroscopic binaries. With the exception of
S\,1040 and S\,1112, all variables have orbital periods less
than 20 days and $V>$\,15.  In all cases but S\,1019 and S\,1063, the
photometric period is equal to the orbital period or, in the case of
S\,1242, the orbital period near periastron.  Evidently tidal locking
has been established, leading to rotation of at least the primary star
that is more rapid than typical for solar-mass stars at 4 Gyr. Thus
our results establish a key premise of the Belloni et al. (1998)
picture for the X-ray emission. Furthermore, if the cause of the
observed optical variability is indeed spot modulation of the observed
flux, then the presence of the required large spots is consistent with
enhanced magnetic activity in these stars.  The X-ray emission and
optical variability properties of S\,1019 and S\,1063 require further
investigation.

Three binaries were not detected as variables despite their short
orbital periods. S\,972 is the faintest of the binary sample at
$V$\,=\,15.37, and so its variability may have gone undetected. The X-ray
luminosities of S\,1045 ($P_{{\rm orb}}$\,=\,7.6 days) and S\,1234 ($P_{{\rm
orb}}$\,=\,4.3 days) are among the lowest of the binary X-ray sources and
indicate low activity levels; this can explain the absence of optical
variability due to spots.  Rajamohan et al. (1998) have noted S\,1234
as a possible optical variable (semi-amplitude $\sim$0.16 mag) which  
suggests that time-variability of the spot phenomenon can also 
explain the absence of optical variation.

The interpretation of S\,1040 and of the remaining three X-ray
binaries S\,760, S\,1072 and S\,1237 may be the most challenging. All
have long orbital periods.  Given their wider separations tidal
locking is not expected  and so the consequent stellar rotations may be
characteristic of single stars. As such, their lack of large spots and
consequent photometric variability is not a surprise.  Nonetheless,
these binaries are X-ray sources. Their X-ray emission remains to be
explained.

No large radial-velocity variations were found for S\,775
and S\,1270 ($\sigma$ is 0.9 km s$^{-1}$ in 12 observations spanning 5200
days and 0.7 km s$^{-1}$ in 7 observations spanning 800 days,
respectively, see Mathieu et al. 1986); if these stars are binaries
their periods must be relatively long. Thus their X-ray luminosities,
as those of S\,364, S\,628 and S\,1027 for which no radial-velocity
information is available, remain unexplained.

\nocite{rajaea} \nocite{kara} \nocite{mathea86}

\begin{acknowledgements}
The authors wish to thank Magiel Janson, Rien Dijkstra, Gertie
Geertsema, Remon Cornelisse and Gijs Nelemans for obtaining part of
the data used in the paper. The authors want to thank David Latham for
computing preliminary orbital solutions for four spectroscropic
binaries to support this research; the radial-velocity measurements
are part of a larger study of M\,67 binaries carried out by
D.W. Latham, A.A.E. Milone and R.D. Mathieu. The Kitt Peak National
Observatory is part of the National Optical Astronomy Observatories,
which is operated by the Association of Universities for Research in
Astronomy, Inc. (AURA) under cooperative agreement with the National
Science Foundation. The Jacobus Kapteyn Telescope is operated on the
island of La Palma by the Isaac Newton Group in the Spanish
Observatorio del Roque de los Muchachos of the Instituto de
Astrofisica de Canarias. The Dutch 0.91m Telescope was operated at La
Silla by the European Southern Observatory. MvdB was supported by the
Netherlands Organization for Scientific Research (NWO).
\end{acknowledgements} 

\nocite{vdbergea2001ad}

\bibliographystyle{apj}

\end{document}